\documentclass{article}
\usepackage{spconf,amsmath,graphicx,hyperref}
\usepackage{amsfonts}
\usepackage{booktabs}
\usepackage{subcaption}

\def\T{{\textsf{T}}}

\title{Improving DF-Conformer using Hydra for high-fidelity \\ generative speech enhancement on discrete codec token}

\name{Shogo Seki$^{*}$, Shaoxiang Dang$^{*}$, Li Li$^{*}$
\thanks{$^{*}$Equal contribution. Work contributed by Shaoxiang during internship.}
}
\address{AI Lab, CyberAgent, Tokyo, Japan}
\begin{document}
%
\maketitle
\begin{abstract}
The Dilated FAVOR Conformer (DF-Conformer) is an efficient variant of the Conformer architecture designed for speech enhancement (SE). It employs fast attention through positive orthogonal random features (FAVOR+) to mitigate the quadratic complexity associated with self-attention, while utilizing dilated convolution to expand the receptive field. This combination results in impressive performance across various SE models. In this paper, we propose replacing FAVOR+ with bidirectional selective structured state-space sequence models to achieve two main objectives: (1) enhancing global sequential modeling by eliminating the approximations inherent in FAVOR+, and (2) maintaining linear complexity relative to the sequence length. Specifically, we utilize Hydra, a bidirectional extension of Mamba, framed within the structured matrix mixer framework. Experiments conducted using a generative SE model on discrete codec tokens, known as Genhancer, demonstrate that the proposed method surpasses the performance of the DF-Conformer.
\end{abstract}
\begin{keywords}
High-fidelity speech enhancement, Genhancer, state-space models (SSMs), Mamba, Hydra
\end{keywords}
\section{Introduction}
\label{sec:intro}
With the rapid advancements in deep learning, speech enhancement (SE) has transcended its traditional boundaries, which primarily focused on isolated tasks such as denoising and dereverberation. The modern broader objective of SE is to generate a high-fidelity version of noisy input, potentially recovering significant missing information. By harnessing the powerful speech generation capabilities of neural vocoders and neural codecs, SE methods \cite{Miipher,miipher-2,reverbmiipher,Genhancer,DiTSE,SELM} can effectively produce high-fidelity speech from the denoised features of degraded inputs.
\par
Genhancer \cite{Genhancer} exemplifies this evolution by employing discrete tokens to achieve remarkable performance and offering significant flexibility for integration with various modalities and techniques in speech processing. It generates clean speech as Descript audio codec (DAC) \cite{DAC} tokens from denoised features, with waveforms reconstructed by a DAC decoder. At the core of Genhancer's feature cleaning and token generation is the dilated FAVOR Conformer (DF-Conformer) \cite{DF-Conformer}, an efficient variant of the Conformer \cite{conformer}. The DF-Conformer employs a macaron-like architecture that incorporates fast attention through positive orthogonal random features (FAVOR+) \cite{performer}, reducing the quadratic complexity of self-attention to linear, alongside dilated convolution (DC) \cite{chen2018encoder} to expand the local receptive field. This efficient combination of global and local modeling makes DF-Conformer well-suited for scaling in large generative SE (GSE) models such as Miipher \cite{Miipher} and Genhancer.
However, advanced analyses of linear attention mechanisms \cite{linear_attention}, including FAVOR+, have identified potential performance degradation compared to softmax-based self-attention. This is particularly evident in aspects such as focus ability, feature diversity \cite{FLatten}, injectivity, and local modeling capability \cite{InLine}, primarily because these mechanisms achieve linear complexity by approximating softmax attention. While FAVOR+ can mitigate approximation errors by increasing the number of random features, this improvement comes at the cost of computational efficiency.
\par
On the other hand, structured state space sequence models (SSMs) \cite{S4}, particularly the selective SSMs known as Mamba \cite{mamba,mamba2}, have emerged as a compelling alternative to self-attention, offering linear complexity without the need for approximation. Recent studies have shown that both SSMs and attention mechanisms can be conceptualized as matrix mixer sequence models. Softmax attention employs a dense, full-rank matrix mixer, while linear attentions approximate this using low-rank matrices and carefully designed kernel functions \cite{mamba2,hydra}. In contrast, SSMs achieve linear complexity by utilizing a semiseparable structured matrix mixer. Building on this concept, Hydra \cite{hydra} extends the semiseparable matrix to a quasiseparable form, enabling a natural and superior bidirectional modeling of Mamba.
\par
In this paper, we first experimentally demonstrate that FAVOR+ within the Genhancer framework suffers from performance limitations due to some of the aspects mentioned above. We then introduce {\it DC-Hydra} to mitigate the limitation by replacing the approximation model, FAVOR+, with Hydra, thereby enhancing Genhancer's performance while maintaining linear complexity.

\section{Genhancer}
\label{sec:genhancer}
\subsection{System overview}

\begin{figure}[th]
\centering
\includegraphics[width=\linewidth]{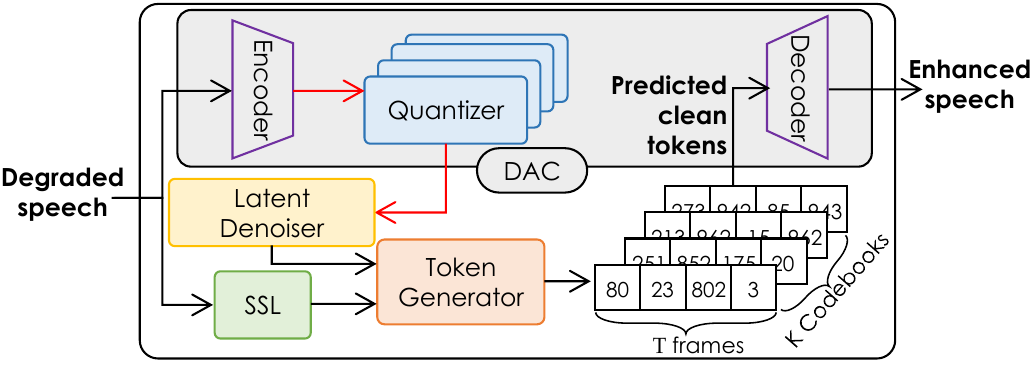}
\vspace{-20pt}
\caption{Overview of Genhancer}
\vspace{-1.2em}
\label{fig:genhancer}
\end{figure}

Genhancer \cite{Genhancer} (Fig.~\ref{fig:genhancer}) utilizes DAC \cite{DAC}, a high-fidelity neural codec comprising an encoder, \(K\) quantizers, and a decoder, to reconstruct clean speech. In DAC, continuous features are represented by \(K\) indices, each corresponding to a \(M\)-dimensional codeword from the respective codebook, with each codebook containing \(I\) codewords. 
Given degraded speech \(\mathbf{x} \in \mathbb{R}^L\), Genhancer reconstructs clean speech \(\hat{\mathbf{y}} \in \mathbb{R}^L\) by estimating the corresponding clean DAC tokens \(\mathbf{Z} \in \mathbb{Z}^{K \times T}\), conditioned on the denoised features \(\mathbf{C} \in \mathbb{R}^{D \times T}\). Here, \(L\), \(T\), and \(D\) denote the sequence length in time domain, the sequence length in feature domain, and the feature dimension, respectively.
The conditional feature $\mathbf{C}$ is primarily derived by denoising feature embeddings extracted from the input using a latent denoiser \(\mathcal{LD}(\cdot)\), given as
$ \tilde{\mathbf{Z}} = {\rm Quantizer}\big({\rm Enc}(\mathbf{x})\big),~ 
\mathbf{C}_{\rm token} = \mathcal{LD}_{\theta_L}\big({\rm Emb}(\tilde{\mathbf{Z}})\big)$.
Here, \(\tilde{\mathbf{Z}} \in\mathbb{Z}^{K \times T} \) represents the noisy DAC tokens, and \(\rm Emb(\cdot)\) denotes the dequantization operation used to retrieve the corresponding codeword from index.
Self-supervised learning (SSL) features can be combined with \(\mathbf{C}_{\rm token}\) to obtain richer feature representation as
$\mathbf{C} = \mathcal{F}_{\theta_F}\big(\mathbf{C}_{\rm token}, {\rm SSL}(\mathbf{x})\big)$,
where \(\mathcal{F}(\cdot)\) denotes a fusion layer, including interpolation to align the two feature sequences and multiple linear layers to project the features to \(D\) dimensions.
Clean tokens \(\mathbf{Z}\) are then estimated autoregressively in a parallel manner using a token generator \(\mathcal{G}(\cdot)\), which takes the previously estimated feature embeddings and the condition \(\mathbf{C}\) as inputs, expressed by
$\mathbf{z}_{k+1} = \mathcal{G}_{\theta_G}\Big(\mathbf{z}_{k+1} \Big|\sum_{k'=0}^k {\rm Emb}(\mathbf{z}_{k'}), \mathbf{C}\Big)$ for 
$k=1,\ldots, K,$
where \({\rm Emb}(\mathbf{z}_0)\) is initialized with zeros. The clean speech is finally reconstructed using a DAC decoder as
$\hat{\mathbf{y}} = {\rm Dec}(\mathbf{Z})$.
In summary, \(\Theta=\{\theta_L, \theta_F, \theta_G\}\) represents the trainable parameters in Genhancer.

\subsection{DF-Conformer}
\label{subsec:dc-conformer}
DF-Conformer \cite{DF-Conformer} serves as the backbone for the latent denoiser \(\mathcal{LD}(\cdot)\) and the token generator \(\mathcal{G}(\cdot)\). It is a Conformer \cite{conformer} variant designed to enhance efficiency by replacing softmax-based self-attention \cite{transformer} with FAVOR+ \cite{performer} and standard convolution with dilated convolution (DC).
\par
Softmax attention utilizes queries, keys, and values, represented as \(\mathbf{Q}, \mathbf{K}, \mathbf{V} \in \mathbb{R}^{T \times d}\), respectively, to transform features using the formula \({\rm SA}(\mathbf{Q}, \mathbf{K}, \mathbf{V}) = {\rm Softmax}(\mathbf{Q}\mathbf{K}^\T)\mathbf{V}\). This operation has a complexity of \(O(T^2d)\), primarily due to the matrix multiplication involved in the softmax function.
FAVOR+ offers an efficient approximation of this transformation with \({\rm FA}(\mathbf{Q}, \mathbf{K}, \mathbf{V}) = \mathbf{D}^{-1}\phi(\mathbf{Q})\big(\phi(\mathbf{K})^\T\mathbf{V}\big)\). 
The approximation is achieved by employing random feature maps \(\phi: \mathbb{R}^d \rightarrow \mathbb{R}^{r}_+\), where \(\mathbf{D}\) is a normalization matrix. With an appropriate number of features \(r\), FAVOR+ can accurately approximate the original softmax attention. By leveraging the approximation and reordering operations, FAVOR+ reduces the complexity to \(O(Trd)\), classifying it as a form of linear attention, which is more scalable for longer sequences.

\subsection{Analysis of FAVOR+}
\begin{figure}[t]
\vspace{-1em}
\begin{subfigure}{\linewidth}
\centering
\includegraphics[width=\linewidth]{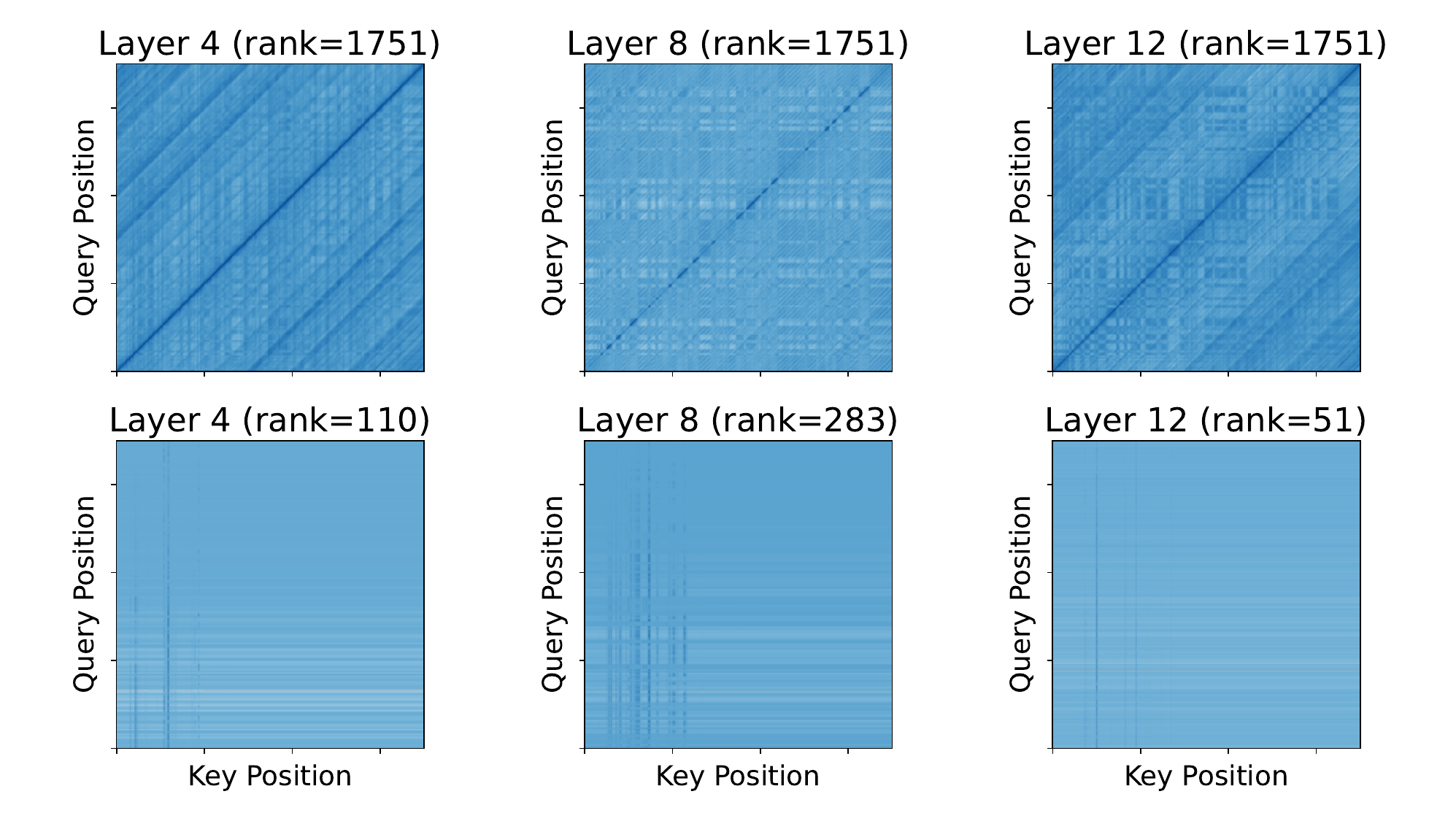}
\end{subfigure}
\begin{subfigure}{\linewidth}
\vspace{-5pt}
\centering
\includegraphics[width=\linewidth]{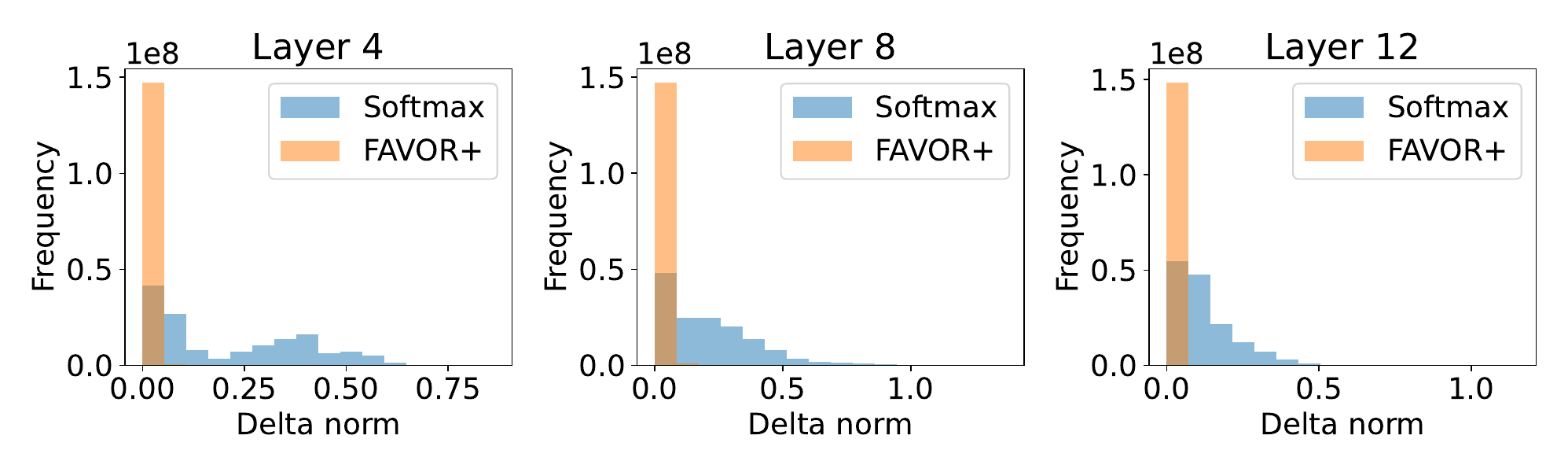}    
\end{subfigure}
\vspace{-20pt}
\caption{Examples of attention maps averaged over heads, along with corresponding ranks in different layers, obtained using softmax attention (1st row) and FAVOR+ (2nd row). Histogram of $L2$ norm difference between attention vectors for different queries (3rd row).}
\vspace{-1.2em}
\label{fig:attention_maps}
\end{figure}

Recent findings \cite{FLatten,InLine} indicate that linear attentions, due to their approximation of softmax attention, can sometimes result in a performance gap. This gap is partly attributed to how linear attentions handle certain key properties:
(a) Focus ability: The capacity to precisely highlight or concentrate on specific, relevant parts of the input.
(b) Feature diversity: The ability to combine a wide variety of useful features from all values.
(c) Injectivity: The capacity for the attention function to be injective, ensuring distinct queries result in distinct attention maps; otherwise semantic confusion occurs.
(d) Local modeling capability: The ability to pay more attention to the neighborhoods of each query in shallow layers.
It is important to note that these properties are not isolated but interconnected. For instance, if the attention function is non-injective, causing different queries to produce identical attention patterns, it can directly result in reduced feature diversity.
\par
Our analysis of FAVOR+ in Genhancer reveals insufficient focus ability, reduced feature diversity, and occurrences of semantic confusion.
In the first and second rows of Figure~\ref{fig:attention_maps}, we visualize the attention maps of FAVOR+ and softmax attention to intuitively assess focus ability and compute matrix ranks to quantify feature diversity. FAVOR+ generates more blurred attention maps with low ranks, whereas softmax attention produces sharp attention patterns (characterized by several deep diagonal lines) with full ranks.
In the third row, we present histograms of the \(L2\) norm differences between attention vectors (each row vector in the attention map). It is observed that FAVOR+ generates similar attention vectors for almost all queries, indicating significant semantic confusion occurs among queries.

\section{Proposed method: DC-Hydra}

SSMs \cite{S4} achieve linear complexity by compressing information from previous frames using hidden states and leveraging recurrence. Mamba \cite{mamba} introduces a selective mechanism that dynamically adjusts sequence modeling parameters based on input, acting as a gating mechanism and achieving performance comparable to Transformers \cite{transformer}. Recently, Mamba-2 \cite{mamba} and its bidirectional extension, Hydra \cite{hydra}, have been developed by reformulating SSMs within a matrix mixer sequence model framework. As a promising alternative for achieving linear complexity in attentions, we propose replacing FAVOR+ with Hydra to address previous limitations. Our module, called {\bf DC-Hydra}, combines dilated convolution (DC) with Hydra for both local and global modeling.

\subsection{Matrix mixer sequence models}
Let $\boldsymbol{X} \in \mathbb{R}^{T \times d}$ be an input sequence. The term {\it sequence transformation} refers to a mapping where the output sequence $\boldsymbol{Y} \in \mathbb{R}^{T \times d}$ can be represented by
$\boldsymbol{M}_\theta = f_\mathcal{M}(\boldsymbol{X}, \theta),~\boldsymbol{Y} = \boldsymbol{M}_\theta\boldsymbol{X}$.
Here, $\boldsymbol{M} \in \mathbb{R}^{T \times T}$ is a matrix mixer, and $f_\mathcal{M}$ denotes a function to generate input-dependent mixer. $\mathcal{M}$ represents the underlying class of mixer matrices and $\theta$ are learnable parameters.
In this content, softmax attention can be interpreted as a dense matrix mixer applied to values $\mathbf{V}$ as the input sequence, where $\boldsymbol{M}_\theta={\rm Softmax}(\mathbf{Q}\mathbf{K}^\T)$. 
FAVOR+ is interpreted as a low-rank matrix mixer with rank of $r$ applied to values $\mathbf{V}$ with $\boldsymbol{M}_\theta = \mathbf{D}^{-1}\phi(\mathbf{Q})\phi(\mathbf{K})^\T$.
\par
Within the matrix mixer framework, both Mamba and Mamba-2 can be expressed as inputs transformed by a semiseparable matrix mixer. Let us recall the original recurrent formula of Mamba, where \(d\)-dimensional features are transformed independently.
\vspace{-0.3\baselineskip}
\begin{align}
\boldsymbol{h}_{t} =  \boldsymbol{A}_t \boldsymbol{h}_{t-1} + \boldsymbol{b}_t x_t,~y_t = \boldsymbol{c}_t^\T \boldsymbol{h}_{t},
\end{align}
where $\boldsymbol{A}_t \in \mathbb{R}^{N \times N} $, $\boldsymbol{b}_t \in \mathbb{R}^{N}$ and $\boldsymbol{c}_t \in \mathbb{R}^{N}$ are time-varying parameters discretized using an input-dependent parameterized step size $\Delta_t$, and $N$ denotes hidden state size.
By expanding the recurrent formula, we can readily derive a matrix multiplication form as
\vspace{-0.5\baselineskip}
\begin{align}
y_t = \sum_{s=0}^t \boldsymbol{c}_t^\T\boldsymbol{A}_{t:s}^{\times}\boldsymbol{b}_sx_s,
~~
\boldsymbol{A}_{i:j}^{\times} 
&= \begin{cases}
 \prod_{k=j+1}^{i} \boldsymbol{A}_k, & \! i > j, \\[6pt]
1, & \! i = j, \\[6pt]
 \prod_{k=i}^{j-1} \boldsymbol{A}_k, & \! i < j .
\end{cases}
\end{align}
The sequence transformation can be represented with a semiseparable matrix mixer, the $ij$-th element of whose is
$m_{ij} = \boldsymbol{c}^\T_i\boldsymbol{A}_i\cdots\boldsymbol{A}_{j+1}\boldsymbol{b}_j$.


\subsection{Hydra and DC-Hydra backbone}
\begin{figure}[t]
\centering
\includegraphics[width=0.7\linewidth]{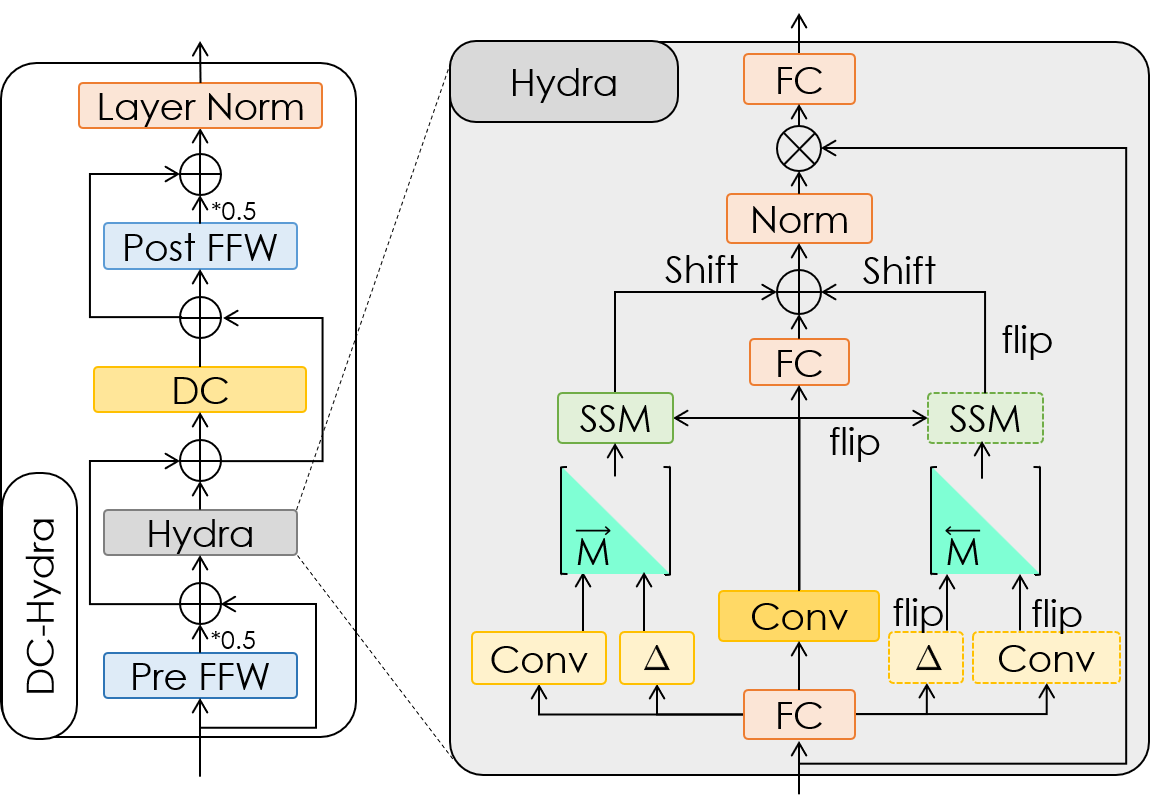}
\vspace{-10pt}
\caption{Architecture of DC-Hydra.}
\vspace{-1.2em}
\label{fig:dchydra}
\end{figure}

To comprehensively explore sequence information, bidirectional variants of Mamba (Bi-Mamba) have been extensively studied \cite{zhu2024vision,schiff2024caduceus,zhang2025mamba,miyazaki2024exploring}.
A straightforward method to achieving bidirectionality involves using two separate Mamba models to handle forward and backward sequence modeling, followed by fusing the knowledge from both models via operation such as addition.
Recently, Hydra \cite{hydra} has been proposed as a mathematical extension of bidirectional Mamba within the matrix mixer framework, where the matrix mixer is defined as a quasiseparable matrix. 
The elements in matrix mixers of addition-based Bi-Mamba and Hydra, represented as $\ddot{m}_{ij}$ and $\breve{m}_{ij}$ respectively, are given as
\vspace{-0.3\baselineskip}
\begin{align*}
\hspace{-0.5em}
\ddot{m}_{ij}
  = \begin{cases}
        \overrightarrow{\boldsymbol{c}_i^\T} \overrightarrow{\boldsymbol{A}_{i:j}^{\times}} \overrightarrow{\boldsymbol{b}_j} , & \!\!\! \! i > j, \\
       \overrightarrow{\boldsymbol{c}_i^\T} \overrightarrow{\boldsymbol{b}_j} + \overleftarrow{\boldsymbol{c}_i^\T} \overleftarrow{\boldsymbol{b}_j}, & \! \!\!\! i = j, \\
      \overleftarrow{\boldsymbol{c}_i^\T} \overleftarrow{\boldsymbol{A}_{i:j}^{\times}} \overleftarrow{\boldsymbol{b}_j} , & \!\!\! \! i < j,
     \end{cases}
     ~~
\breve{m}_{ij}
  = \begin{cases}
        \overrightarrow{\boldsymbol{c}_{i-1}^\T} \overrightarrow{\boldsymbol{A}_{i-1:j}^{\times}} \overrightarrow{\boldsymbol{b}_j} , & \!\! \! i > j, \\
       \delta_i, & \!\! \! i = j, \\
      \overleftarrow{\boldsymbol{c}_{i+1}^\T} \overleftarrow{\boldsymbol{A}_{i+1:j}^{\times}} \overleftarrow{\boldsymbol{b}_j} , & \!\! \! i < j.
     \end{cases}
\end{align*}
The key difference is that, in addition-based Bi-Mamba, diagonal elements are influenced by shared non-diagonal parameters, whereas Hydra models them separately, providing stronger representation power.

Fig.~\ref{fig:dchydra} shows the architecture of the proposed DC-Hydra backbone and the implementation details of Hydra. The Hydra and depthwise convolution with dilation (DW Conv) modules are sandwiched between two feed-forward (FFW) modules. Residual connections are applied to all modules, and LayerNorm is applied to the output. The official Hydra implementation\footnote{\url{https://github.com/goombalab/hydra}} using Mamba-2 as the SSM is utilized in the DC-Hydra.

\section{Experiments}
\label{sec:Experiments}

\subsection{Dataset}
\label{ssec:Dataset}

Following~\cite{Genhancer}, we used public speech, noise, and impulse response data to train Genhancer models.
We used speech samples from LibriTTS-R \cite{libritts-r}, where each utterance was upsampled to 48~kHz with a distributed bandwidth extension model~\cite{bandwidth}\footnote{\url{https://github.com/brentspell/hifi-gan-bwe}} and resampled at 44.1~kHz to meet the DAC in-out.
We used noise data from the TAU Urban Audio-Visual Scenes 2021~\cite{urban}, DNS Challenge~\cite{DNSChallenge}, and SFS-Static~\cite{SFS} datasets, and impulse response data from the MIT IR Survey~\cite{traer2016statistics}, EchoThief~\cite{Echothief}, and OpenSLR28~\cite{ko2017study}.
Degraded speech was generated in an on-the-fly fashion by convolving an impulse response and superimposing one or two noise samples with signal-to-noise ratios (SNRs) of [-10, 20] dB.
We randomly applied multiple equalizations across five frequency bands and a bandwidth limitation.
For evaluation, we used the DAPS dataset~\cite{DAPS}, which contains studio-quality, minute-long audio clips from 10 male and female speakers recorded in twelve real-world environments. Each speaker read five scripts, resulting in 1200 test samples.

\subsection{Experimental settings and evaluation metrics}

We utilized a distributed 44.1 kHz DAC variant with nine quantizers, each comprising 1024 codewords, which produces 8-dimensional tokens ($K=9$, $M=8$, $I=1024$) at a frame rate of 86~Hz.
For the SSL feature extractor $\mathrm{SSL}(\cdot)$, we used a pre-trained large WavLM model\footnote{\url{https://huggingface.co/microsoft/wavlm-large}}.
The intermediate layer outputs of the WavLM model were combined using learnable weights to generate SSL features.
The latent denoiser $\mathcal{LD}(\cdot)$ and the token generator $\mathcal{G}(\cdot)$ employed DF-Conformer blocks with 256 and 512-channels , consisting of 8 and 12 blocks, respectively.
Additionally, the convolution kernels were dilated by a scale factor of 2 every four blocks.

We investigated the following four module alternatives to attention in DF-Conformer winthin Genhancer.
{\bf FAVOR+:} The baseline Genhancer model with 98M parameters.
{\bf Softmax:} Similar to FAVOR+ with 98 million parameters, but utilizing softmax attentions instead of FAVOR+ attentions.
{\bf Bi-Mamba:} A drop-in replacement using Mamba blocks with forward and backward SSMs~\cite{zhu2024vision} (107 M parameters).
{\bf Hydra:} The proposed Genhancer model, consisting of bidirectional SSM blocks, with 106 million parameters.
For FAVOR+ and Softmax, rotary position embeddings~\cite{su2024roformer} were applied to keys and queries at each DF-Conformer block. All models were trained for 400,000 steps using 8-second input and minibatches of size 16 on four NVIDIA A100 GPUs, taking approximately five days. We used the AdamW optimizer with a cosine learning rate scheduler, including a warmup period. The learning rate was initially increased linearly from $1\mathrm{e}{-5}$ to $1\mathrm{e}{-4}$ over the first 1,000 steps, then decreased back to $1\mathrm{e}{-5}$ following a cosine curve over 300,000 steps. During inference, the input speech was enhanced by dividing it into 8-second segments, identical to the training size, and merging these chunks. We also examined the effects of sequence length differences between training and inference.

We used a well-known GSE model Miipher~\cite{Miipher} as a reference.
An open-source model\footnote{\url{https://github.com/Wataru-Nakata/miipher.git}} with 105 M parameters was trained with the same datasets.
We utilized two non-intrusive SE metrics: DNSMOS~\cite{reddy2021dnsmos} and UTMOS~\cite{saeki2022utmos} to assess enhanced speech quality, and speaker similarity (SpkSim)~\cite{jung24c_interspeech} to evaluate speaker consistency.  
Additionally, we employed character accuracy (CAcc) using an Open Whisper-style speech model (OWSM) \cite{peng2024owsm} to measure content accuracy. 

\subsection{Results}

\begin{table}
\footnotesize
\caption{Mean DNSMOS, UTMOS, and speaker similarity (SpkSim) scores and character accuracy (CAcc), where bold fonts indicate the best performance between models.}
\label{tab:result}
\centering
\begin{tabular}{lrrrr}\toprule
                            & DNSMOS$\uparrow$ & UTMOS$\uparrow$ & SpkSim$\uparrow$ & CAcc [\%]$\uparrow$ \\ \midrule     
     Clean                  & 3.39             & 3.83            & N/A              & 91.35               \\              
     Noisy                  & 2.56             & 1.70            & 0.91             & 90.93               \\ \midrule     
     Miipher~\cite{Miipher} & 3.33             & 2.77            & 0.73             & 87.82                 \\
     Softmax                & \textbf{3.46}    & \textbf{3.53}   & \textbf{0.83}    & 87.88               \\              
     FAVOR+                 & 3.44             & 3.33            & 0.79             & 88.24               \\              
     Bi-Mamba               & 3.44             & 3.27            & 0.81             & 88.04               \\              
     Hydra (ours)           & 3.44             & 3.48            & \textbf{0.83}    & \textbf{88.95}      \\ \bottomrule  
\end{tabular}
\normalsize
\end{table}

\begin{figure}[t]
    \includegraphics[width=\linewidth]{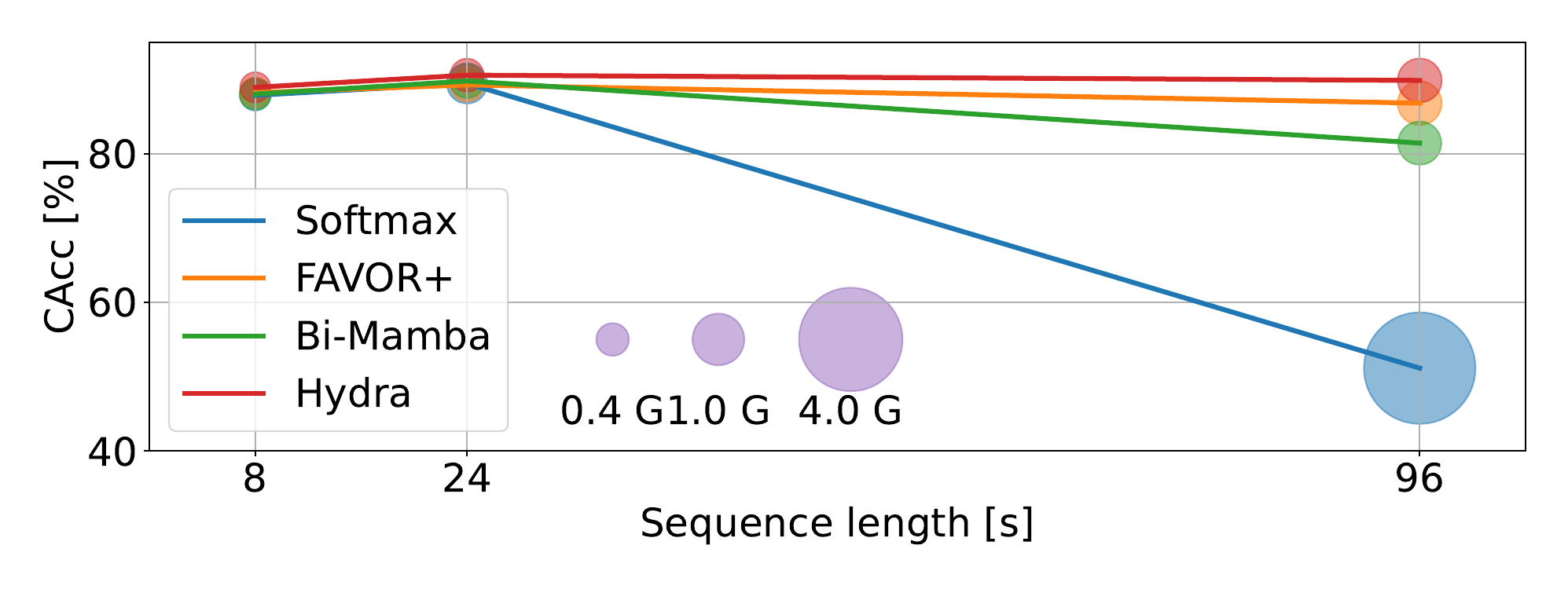}
    \vspace{-20pt}
    \caption{Character accuracy (CAccs) on different sequence lengths, with the babble size indicating GPU memory usage in the token generator $\mathcal{G}$.}
    \label{fig:length}
    \vspace{-1.2em}
\end{figure}

Table~\ref{tab:result} compares Genhancers with various DF-Conformer modules. Genhancer-based models outperform Miipher, demonstrating superior performance. The Softmax model, with its quadratic complexity, achieves the highest scores in DNSMOS, UTMOS, and SpkSim, serving as the upper bound. Among the remaining methods, Hydra outperforms both FAVOR+ and Bi-Mamba and even surpasses Softmax in CAcc.

Fig.~\ref{fig:length} compares models with varying input lengths. While the SSL feature extractor primarily drives the computational cost of Genhancer, GPU memory usage in the token generator $\mathcal{G}$ is also included for comparison. Models handling 24-second inputs perform similarly to those with 8-second inputs. However, for 96-second inputs, performance degradation is evident, notably with a significant drop in the Softmax model. Interestingly, compared to the Softmax model, the baseline FAVOR+ maintains performance well for longer inputs. This may be due to FAVOR+'s difficulty in approximating softmax attention, as shown in Fig.~\ref{fig:attention_maps}, contributing minimally to sequential modeling and thus minimizing the impact of longer sequence lengths. The proposed Hydra-based Genhancer excels among these models, showcasing Hydra's effectiveness in GSE.

\section{Conclusions}
In this paper, we analyzed FAVOR+, a crucial component of DF-Conformer used in Genhancer. Our analysis revealed that FAVOR+ suffers from low focus ability, reduced feature diversity, and semantic confusion, similar to other linear attention mechanisms, leading to performance limitations. To address this, we proposed DC-Hydra, which replaces FAVOR+ with a mathematically extended bidirectional Mamba, enhancing the SE performance of Genhancer while maintaining linear complexity in sequence modeling. Experimental results confirmed the effectiveness of DC-Hydra.

\newpage
\footnotesize
\bibliographystyle{IEEEbib}
\bibliography{strings,refs}

\end{document}